\documentstyle[12pt]{article}
\input tcilatex

\begin{document}

QUALITATIVE ANALYSIS ON THE EXISTENCE OF BOUND EXCITED STATES AND LOW-LYING
RESONANCES OF THE DIPOSITRONIUMS

\hspace{1.0in}

C.G.Bao

Department of Physics, Zhongshan University, Guangzhou, 510275, China

\hspace{1.0in}

ABSTRACT: Symmetry has imposed very strong constraints on the structure of
the internal wavefunctions and on the accessibility of outgoing channels.
Based on symmetry consideration, the existence of a number of bound excited
states and low-lying resonances has been suggested.

\hspace{1.0in}


\hspace{1.0in}

\hspace{1.0in}

The existence of the dipositroniums Ps$_2$ has not yet been experimentally
confirmed, but has already been predicted by a number of theoretical
calculations [1-6]. Recently, in addition to the ground state, a bound
excited state with total orbital angular momentum L=1 and parity $\Pi =-1$
has also been predicted [7,8]. How many bound states will this system
contain? What is the feature of the low-lying spectrum? These questions are
attractive. In this paper we shall study this problem based on symmetry
consideration and based on existing theoretical results.

Since the Hamiltonian of the dipositronium is invariant to the permutations
isomorphic to the point group D$_{2d}$ [2], The eigenstates can be
classified according to the representations $\mu $ of the D$_{2d}$ group, $%
\mu =A_1,A_2,B_1,B_2,$ or $E..$ Since the Hamiltonian is also invariant to
rotation and space inversion, an eigenstate can be labeled as L$_i^\Pi (\mu
),$ $i$ is a serial number for a series of states with the same (L,$\Pi $, $%
\mu )$ $[9].$ When $i=1$, it is the lowest state of this series; in this
case the label $i$ will be omitted. It is noted that the spin-state is
determined by $\mu $. Let the two electrons be the particles 1 and 2, the
positrons be 3 and 4. Let the spins of 1 and 2 be coupled to s$_1,$ and the
spins of 3 and 4 be coupled to s$_2$. When $\mu =A_1$ or $B_2,$ we have ($%
s_1,s_2)=(0,0);$when $\mu =B_1$ or $A_2,$ we have ($s_1,s_2)=(1,1);$when $%
\mu =E,$ we have ($s_1,s_2)=(0,1)$ or (1,0).

Evidently, the stability of an eigenstate depends on the coupling of this
state with the open channels. Whether a channel is open or closed is
determined not only by energy but also by symmetry. For the dipositronium,
the most important (the lowest) outgoing channel is the Ps-Ps channel ,
where both Ps are in the ground state, with threshold energy $-0.5$ (atomic
units are used in this paper). Let us see how the symmetry will impose
constraints on this channel.

(i) Let $l$ be the relative orbital angular momentum between the two
positroniums. Since each positronium is in its ground state with zero
orbital angular momentum, we have $l=$L and parity $\Pi =(-1)^L$. This fact
implies that for all the states with $\Pi (-1)^L=-1,$ the Ps-Ps channel is
closed.

(ii) Let $p_{ij}$ denotes an interchange of i and j. Since s$_1$ and s$_2$
are good quantum numbers, an eigenstate is written as

$\Psi =[(1+(-1)^{s_1}p_{12})(1+(-1)^{s_2}p_{34})F]\chi _{s_1s_2}$%
\hspace{1.0in}(1)

where $F$ is a function of the spatial coordinates and $\chi _{s_1s_2}$ is
the spin state. In the Ps-Ps channel, the spatial wavefunction can be
written as

$F_{open}=(1+(-1)^{s_1}p_{12})(1+(-1)^{s_2}p_{34})\Phi (\stackrel{%
\rightarrow }{r}_{13})\Phi (\stackrel{\rightarrow }{r}_{24})f_L(\stackrel{%
\rightarrow }{r}_{13,24})$

$=(1+(-1)^{s_1+s_2+L})[\Phi (\stackrel{\rightarrow }{r}_{13})\Phi (\stackrel{%
\rightarrow }{r}_{24})f_L(\stackrel{\rightarrow }{r}_{13,24})+(-1)^{s_1}\Phi
(\stackrel{\rightarrow }{r}_{23})\Phi (\stackrel{\rightarrow }{r}_{14})f_L(%
\stackrel{\rightarrow }{r}_{23,14})]$\hspace{1.0in}(2)

where $\stackrel{\rightarrow }{r}_{13}=\stackrel{\rightarrow }{r}_3-%
\stackrel{\rightarrow }{r}_1$, $\stackrel{\rightarrow }{r}_{13,24}=\frac 12(%
\stackrel{\rightarrow }{r}_2+\stackrel{\rightarrow }{r}_4)-\frac 12(%
\stackrel{\rightarrow }{r}_1+\stackrel{\rightarrow }{r}_3),$ etc; and $%
\stackrel{\rightarrow }{r_i}$ is the position vector of the i-th particle
originating from the c.m.. $\Phi $ denotes the ground state of a
positronium; $f_L$ is for the relative motion of the two positroniums with
angular momentum $L$. From (2) it is obvious that for all the states with $%
s_1+s_2+L=odd$, the Ps-Ps channel is closed.

(iii) When ($s_1,s_2)=(0,0)$ or (1,1), the spatial wavefunction is an
eigenstate of $p_{13}p_{24}$ with the eigenvalues $\Lambda =\pm 1$[2,9]$.$
In this case the $F_{open}$ is nonzero only if L=even. When ($s_1,s_2)=(0,0)$
or (1,1) and L=even, one can prove that

$p_{13}p_{24}F_{open}=F_{open}$\hspace{1.0in}(3)

Eq.(3) implies that, for all the states with $\Lambda =-1,$ the Ps-Ps
channel is closed.

Eq.(1) to (3) impose a strong constraint on the quantum numbers (L, $\Pi ,$ $%
\mu )$ of the open channels. Let us consider all the L$\leq 2$ states (the
discussion can be generalized to L$\geq 3$ states as well). There are thirty
types of (L, $\Pi ,$ $\mu )$ symmetries involved as listed in Table 1.
However, owing to the above constraints, only the

0$_i^{+}(A_1),$0$_i^{+}(B_1),$1$_i^{-}(E),$2$_i^{+}(A_1)$, and 2$_i^{+}(B_1)$

states are allowed to access the Ps-Ps channel, while all the states of the
other twenty-five symmetries are not allowed. The accessibility of the Ps-Ps
channel is listed in Table 1. Since the Ps-Ps channel is closed to most L$%
^\Pi (\mu )$ states, the existence of a number of bound states is possible.
In fact, for the states with the Ps-Ps channel closed, those with their
energies lower than the second lowest threshold Ps-Ps* (one of the
positronium is at the first excited states) are definitely bound .

The above procedure can be generalized to analyses any out going channel.
For example, let us inspect the Ps-Ps*(01) channel, in which the excited
positronium has ($n_2l_2)=(01).$ In this case, we have

$F_{open}=(1+(-1)^{s_1}p_{12})(1+(-1)^{s_2}p_{34})\stackunder{l}{\sum }%
\{\Phi (\stackrel{\rightarrow }{r}_{13})\Phi _{01}(\stackrel{\rightarrow }{r}%
_{24})f_l(\stackrel{\rightarrow }{r}_{13,24})\}_L$\hspace{1.0in}(4)

where $\Phi _{01}$ is the internal state of the excited positronium, $l$ is
the relative angular momentum. It can be rewritten as

$F_{open}$=$\stackunder{l}{\sum }\{[\Phi (\stackrel{\rightarrow }{r}%
_{13})\Phi _{01}(\stackrel{\rightarrow }{r}_{24})+(-1)^{s_1+s_2+l}\Phi (%
\stackrel{\rightarrow }{r}_{24})\Phi _{01}(\stackrel{\rightarrow }{r}%
_{13})]f_l(\stackrel{\rightarrow }{r}_{13,24})$

$+(-1)^{s_1}[\Phi (\stackrel{\rightarrow }{r}_{23})\Phi _{01}(\stackrel{%
\rightarrow }{r}_{14})+(-1)^{s_1+s_2+l}\Phi (\stackrel{\rightarrow }{r}%
_{14})\Phi _{01}(\stackrel{\rightarrow }{r}_{23})]f_l(\stackrel{\rightarrow 
}{r}_{23,14})\}_L$\hspace{1.0in}(5)

When ($s_1,s_2)=(0,0)$ or (1,1) , we have

$p_{13}p_{24}F_{open}=-F_{open}$\hspace{1.0in}(6)

Therefore, this channel is closed if $\Lambda =+1.$ Furthermore, if L=0, we
have $l=1$ and $\Pi =+1;$ therefore this channel is closed for L$^\Pi =0^{-}$
states. The analysis on the accessibility of outgoing channels is summarized
in the last three columns of Table 1.

Another important factor affecting the low-lying spectrum is the
accessibility of specific geometric configurations. Let $A$ denote a
geometric configuration, let $\Delta $ denote a comprehensive symmetric
operation (successive operations of rotation, permutation, and space
inversion). In some cases we have $\Delta A=A.$ For example, if $A$ is a
square with the two electrons 1 and 2 at the two ends of a diagonal and 3
and 4 at the two ends of the other diagonal (denoted as SQ hereafter), then $%
A$ is invariant to a space inversion together with $p_{12}p_{34}.$ $A$ is
also invariant to a rotation about the normal of the plane by 90$^{\circ }$
together with a cyclic permutation of particles, and invariant to a rotation
about a diagonal by 180$^{\circ }$ together with an interchange of the two
particles at the other diagonal. Let the operator $\Delta ^{\prime }$ be
defined as $\Delta ^{\prime }\Psi (A)=\Psi (\Delta A).$ When $A$ is
invariant to $\Delta $, we have

($\Delta ^{\prime }-1)\Psi (A)=0$\hspace{1.0in}(7)

Owing to the transformation property inhering in $\Psi $ under symmetric
operations, (7) is equivalent to a set of homogeneous linear algebra
equations that $\Psi $ has to fulfill at $A$. The details of the equations
depend on the transformation property of $\Psi $ , i.e., depend in general
on (L, $\Pi ,$ $\mu )$ . It is well known that homogeneous equations do not
always have nonzero solution. In some cases, there are only zero solutions.
In this case $\Psi $ has to be zero at $A$ , i.e., there is a node at $A$
and $\Psi $ is prohibited to access the shape $A$. For example, when $A$ is
the SQ and $\Delta $ is a space inversion together with $p_{12}p_{34}$, (7)
is rewritten as

((-1)$^{s_1+s_2}\Pi -1)\Psi (A)=0$\hspace{1.0in}(8)

Thus, for all the states with (-1)$^{s_1+s_2}\Pi =-1,\Psi $ must be zero at
the SQ disregarding the size and orientation of the SQ. Therefore these
states are prohibited to access the SQ.

The identification of the accessibility has been given in detail in [9]. The
results are summarized here in Table 1. When the particles form a geometric
configuration in which the repulsive interaction can be overcome by
attractive interaction (e.g., the SQ), then the domain in the
multi-dimensional coordinate space surrounding this configuration is
favorable to binding and would be preferred by the low-lying states. On the
other hand, unfavorable configurations are not important to low-lying
states, these configurations are not included in Table 1. It is noted that
eq.(4) is irrelevant to the orientation and the size of the shape $A$ [9].
Therefore, if $\Psi $ can not access $A$, it can not access all the shapes
which are different from $A$ by size or by orientation. It leads to a fact
that inherent nodal surface would exist in the wavefunctions at the shapes
that can not be accessed. It is well known that the low-lying states do not
prefer to contain nodal surfaces. The more nodal surfaces are contained, the
higher the energy. For example, for the positronium, the ground state does
not contain nodal surfaces. The degenerate first excited state either
contains a node associated with radial motion (if L=0), or contains a node
associated with angular motion (if L=1); the second excited state contains
two nodal surfaces, etc. Therefore, the inaccessibility of shapes implies
that inherent nodal structure has been imposed on the wavefunctions by
symmetry, this inherent nodal structure will affect the energy seriously.
For the states prohibited to access many shapes, many nodal surfaces are
contained, thus these states must be high in energy. On the other hand,
there are some states do not contain inherent nodal surfaces, namely the 0$%
^{+}(A_1),$1$^{-}(E),$2$^{+}(A_1),$ and 2$^{+}(B_1)$ , their wavefunctions
can be optimized to lower the internal energy without suffering restriction.
Therefore these states, if they are bound, must be lower.

Based on the knowledge provided by Table 1and based on the existing
theoretical results, we can deduce the existence of a number of excited
bound states and the feature of the low-lying spectrum as follows.

Besides the ground state 0$_{}^{+}(A_1)$ which has already been predicted to
be bound by many authors [1-6], Kinghorn and Poshusta have calculated all
the 0$^{+}$ states [2]. For the 0$^{+}(E)$ and 0$^{+}(B_2)$, the upper
limits of energy are -0.33 and -0.3145 , respectively. Both values are lower
than the Ps-Ps* threshold at -0.3125. From Table 1 we know that they can not
access the Ps-Ps channel. Thus, these two states should be bound. For the 0$%
^{+}(E)$ state, the average separation between the electrons $\langle
r_{12}\rangle =9.56$, while the average separation between an electron and a
positron $\langle r_{13}\rangle =8.34$ . For the 0$^{+}(B_2)$ state, $%
\langle r_{12}\rangle =12.6$ and $\langle r_{13}\rangle =10.2$. The sizes of
these two states are comparable with the ground state (with $\langle
r_{12}\rangle =6.03$ and $\langle r_{13}\rangle =4.48$) , this is a further
evidence that they are bound. Otherwise, if they are resonances, they should
have a much larger size. Between these two the size of the 0$^{+}(B_2)$ is a
little larger than that of the 0$^{+}(E)$, because the 0$^{+}(B_2)$ is quite
close to the Ps-Ps* threshold. For the 0$^{+}(B_1)$ and 0$_2^{+}(A_1)$
states, the upper limits given in [2] are -0.4994 and -0.4995, and the $%
\langle r_{12}\rangle $ are 50.1 and 84.2, respectively. If they really have
their energies a little above the Ps-Ps threshold at -0.5, they are
resonances because the Ps-Ps channel is open to them. However, if their
actual energies are a little lower than the threshold, then they are bound.
For the 0$^{+}(A_2)$ state, the upper limit given in [2] is -0.3121, and the 
$\langle r_{12}\rangle $ is106.1. If this state really has its energy a
little above the Ps-Ps* threshold, it is a resonance because the Ps-Ps*
channel is open. However, if its energy is actually a little lower than the
threshold at-0.3125, then it is bound because the Ps-Ps channel at -0.5 is
closed (cf. Table 1). Thus, a very accurate calculation on the energies of
the 0$^{+}(B_1)$ , 0$_2^{+}(A_1),$ and 0$^{+}(A_2)$ is necessary to identify
whether these states are bound.

From Table 1 it is clear that both the Ps-Ps and Ps-Ps* channels are closed
to the 0$^{-}(\mu )$ states. Besides, by using the same procedure of
analysis as above, one can prove that neither the Ps-Ps*($nl)$ channels
(where one of the positronium is excited to a state with arbitrary $n$ and $%
l)$, nor the Ps$^{-}-e^{+}$ (or Ps$^{+}-e^{-})$ channel with the threshold
at -0.262, nor the Ps$-e^{+}-e^{-}$ channel with the threshold energy -0.25
are open to the 0$^{-}(\mu $) states. This fact can be illustrated simply by
a partial-wave decomposition. Let $l_1,l_2,$ and $l_3$ be the partial waves
associated with the three Jacobian coordinates of a four-body system. Let $%
(l_1l_2l_3)_L$ denotes a partial-wave component of the spatial wave
function, where the angular momenta are coupled to L. It is obvious that for
L=0 odd-parity states, none of the $l_i$ can be zero. If it does, say, $%
l_1=0;$ then $l_2$ must be equal to $l_3$ to assure L=0. This choice results
in having an even parity, and therefore can not be realized in odd-parity
states. However, in the Ps-Ps*$(nl)$ channels and in the Ps$-e^{+}-e^{-}$
channel, the orbital angular momentum of the Ps is zero; in the Ps$%
^{-}-e^{+} $ channel the relative angular momentum between the Ps$^{-}$ and
the $e^{+}$ is zero (because the ground state of Ps$^{-}$ has total orbital
angular momentum zero). Therefore all these channels are closed to the 0$%
^{-}(\mu )$ states. Hence, if any of the 0$^{-}(\mu )$ state is lower than
-0.25, this state is bound because no outgoing channels are available. Thus
the existence of bound 0$^{-}(\mu )$ states is possible. From Table 1 it is
clear that all the 0$^{-}(\mu )$ states contain a number of inherent nodal
surfaces, therefore the energies of these states are high. However, the 0$%
^{+}(A_2)$ contains also a number of inherent nodal surfaces, but the upper
limit of this state is only -0.3121. There is no reason to assert that the 0$%
^{-}(\mu )$ states should be much higher than the 0$^{+}(A_2)$. Thus, it is
very possible that one (or even more than one) bound 0$^{-}(\mu )$ state(s)
would exist in high energy region of the spectrum. The first candidate would
be the 0$^{-}(B_1)$ (due to containing relatively less inherent nodal
surfaces). This suggestion remains to be confirmed. The anticipated bound
states are listed in Table 2

Usukura, Varga, and Suzuki have calculated a 1$^{-}$ state with ($%
s_1,s_2)=(0,0)$ [8]. The $\mu $ of this state has two possibilities $A_1$ or 
$B_2$. However, since the 1$^{-}(A_1)$ contains a number of inherent nodal
surfaces as shown in Table 1, it should be much higher than the 1$^{-}(B_2)$
. Therefore, the state calculated in [8] with an energy -0.3344 should be
the 1$^{-}(B_2)$ . This state is coupled with the ground state via the E1
transition, therefore it can be induced via $\gamma -$absorption. Since the
Ps-Ps channel is close to it (cf. Table 1), it is a bound state as first
pointed out in [8]. From Table 1 we know that this state can access the
straight chain CH, it is a shape favorable to binding in which each pair of
repulsive interaction can be overcome by attractive interaction. From Table
1 of [9] we know that this chain prefer to be normal to the direction of L.
The correlation functions given in Fig.3 of [8] confirms that the particles
prefer to be lying in the X-Y plane if L is given along the Z-axis.
Furthermore, the ratio $\langle r_{12}\rangle /\langle r_{13}\rangle $ given
in [8] is 1.17. This ratio for the 0$^{+}(E)$ given in [2] is 1.15. From
Table 1 both the 1$^{-}(B_2)$ and 0$^{+}(E)$ can access the favorable
straight chain denoted as CH in the Table. Therefore, the internal
structures of these two states may be more or less similar. This point
remains to be checked.

From Table 1 we know that the 1$^{-}(A_2)$ can also access the straight
chain . Therefore it is anticipated that the energies of the 1$^{-}(A_2\not
) $ and 1$^{-}(B_2)$ are close. In particular, both would prefer the
chain-structure. Since the Ps-Ps channel is also closed to the 1$^{-}(A_2),$
it is also a bound state. Nonetheless, the 1$^{-}(A_2)$ does not coupled
with the ground state by the E1 transition, therefore it is more difficult
to be observed experimentally.

The third candidate of the bound L=1 states is the 1$^{+}(A_2),$ which is
also not allowed to access the Ps-Ps channel. Although this state can not
access the CH, but it can access the SQ, which is also a shape favorable to
binding. It is anticipated that the 1$^{+}(A_2)$ would have an energy not
much higher than those of the 1$^{-}(B_2\not )$ and 1$^{-}(A_2),$ thereby it
would also be bound.

From Table 1 it is clear that the lowest L=1 state should be the 1$^{-}(E).$
This state, just as the ground state 0$^{+}(A_1),$ can access both the most
favorable configurations SQ and CH. Thus both states are inherent nodeless
and their internal wavefunctions can be optimized free of constraints. Let
the energy be divided as a sum of internal energy and collective rotation
energy E$_{rot}.$ If the 1$^{-}(E)$ is bound, it is reasonable to assume
that its internal energy is close to or a little higher than the ground
state. Accordingly, the excitation energy of the 1$^{-}(E)$ is mainly equal
to the collective rotation energy

E$_{rot}=\frac 1{2I_0}L(L+1)$\hspace{1.0in}(9)

where I$_0$ is the moment of inertia. Let I$_0=$ $4m_er_0^2.$ It is found
that if $r_0=3.95,$ then E$_{rot}=0.016.$ Thus, if $r_0$ is a little and
sufficiently larger than 3.95, the energy of the 1$^{-}(E)$ would be lower
than the Ps-Ps threshold at -0.5, and therefore it is bound. Since the
radius 3.95 is only a little larger than that of the ground state, since an
excited state usually has a larger size, the 1$^{-}(E)$ is possible to be
bound. If it is not bound, it would be a resonance a little higher than the
Ps-Ps threshold emerging during Ps-Ps collisions. Nonetheless, since the
width of a resonance depends on the details of dynamics, and can not be
foretold simply from symmetry consideration. Therefore it is not clear
whether the width is narrow enough so that the above proposed resonance can
be experimentally observed.

Among the L=1 states the 1$^{+}(A_1)$ and 1$^{+}(B_1)$ are the only two that
can not access both the Ps-Ps and Ps-Ps*(01 or 10) channels. Therefore, if
they have an energy lower than the Ps-Ps*(02) channel at -0.2778, they are
bound. From Table 1 we know that they contain many inherent nodal surfaces,
just as the 0$^{+}(A_2)$ does; therefore their internal energies would not
be much higher than the 0$^{+}(A_2)$. Since the upper limit of the energy of
the 0$^{+}(A_2)$ is -0.3121, since the E$_{rot}$ is in general small (it is
not larger than 0.016 if $r_0$is not smaller than 3.95), the 1$^{+}(A_1)$
and 1$^{+}(B_1)$ are anticipated to be bound.

It is noted that the quantum number $\Lambda $ [2,9] would change its sign
during an electric transition. Therefore, the ground state ($\mu =A_1)$
would be coupled with the L$^\Pi (B_2)$ states with $\Pi =(-1)^L$ via
electric transitions. Thus, the 2$^{+}(B_2)$ would be coupled with the
ground state by E2 transition. Let us compare the 2$^{+}(B_2)$ with the 1$%
^{-}(B_2),$ the Ps-Ps channel is closed to both states. It is shown in Table
1 that the 2$^{+}(B_2)$ can access the SQ but not the CH, while the 1$%
^{-}(B_2)$ can access the CH but not the SQ. Therefore the difference in
internal energy between them is not anticipated to be large. The 2$^{+}(B_2)$
would be higher than the 1$^{-}(B_2)$ mainly due to having a larger E$%
_{rot}. $ It is recalled that the 1$^{-}(B_2)$ is lower than the Ps-Ps*
threshold by 0.022. If the 2$^{+}(B_2)$ is not higher than the 1$^{-}(B_2)$
by an amount larger than 0.022, it is bound. Otherwise, it is a resonance
emerging in Ps-Ps*(01) collision and in $\gamma -absorption$.

It is noted that the 2$^{+}(E)$ and the 1$^{-}(B_2).$have similar inherent
nodal structures, the 2$^{-}(E)$ and the 2$^{+}(B_2).$have similar inherent
nodal structures. The internal energies of these states are more or less
close to each other. Thus, the above three L=2 states are either bound
states or resonances close to the Ps-Ps* threshold as listed in Table 2.

The 2$^{-}(A_1)$ and 2$^{-}(B_1)$ are the only two L=2 states which can not
escape from both the Ps-Ps and Ps-Ps*(01 or 10) channels. Although they
contain many inherent nodal surfaces, their internal energies are not
anticipated to be much higher than the 0$^{+}(A_2)$ at -0.3121 [2].
Therefore, it is anticipated that they would have an energy below the
Ps-Ps*(02) threshold at -0.2778, in this case they are bound.

The anticipated bound states and low-lying resonances are summarized in
Table 2. It is noted that the experimental observation of the aove proposed
resonances may be difficult, it depends on the widths.

In summary, based on symmetry consideration, the accessibility of outgoing
channels and a number of important geometric configurations has been
studied. Thereby a number of bound states and a few low-lying resonances
have been anticipated. It was found that the outgoing channels are seriously
constrained by symmetry, the lowest channels are open to only a few L$^\Pi
(\mu )$ states. This is a distinguished feature of the dipositroniums not
existing in many other systems (e.g., not exist in the Ps$^{-}$ system).
Accordingly, a number of bound excited states may exist. Evidently, the
above qualitative results should be checked by accurate theoretical
calculations and by experimental data. Incidentally, the values that the
upper limits of the 0$^{+}(B_1)$ and the 0$^{+}(A_2)$ are -0.4994 and
-0.3121 as given in [2] are important to the above analysis. If these values
are changed, some of the above results have to be changed.

The procedure proposed in this paper can be generalized to study the
qualitative feature of the low-lying spectrum of any few-body system [10].

\hspace{1.0in}

REFERENCES

1, E.A. Hylleraas and A.Ore, Phys. Rev. 71 (1947) 493

2, D.B.Kinghorn and R.D.Poshusta, Phys. Rev. A47 (1993) 3671

3, P.M.Kozlowski and L.Adamowicz, Phys. Rev. A48 (1993) 1903

4, K.Varga and Y.Suzuki, Phys. Rev.C52 (1995) 2885: Phys. Rev. A53 (1996)
1907

5, A.M.Frolov and V.H.Smith, Jr., J. Phys. B29 (1996) L433; Phys. Rev. A55
(1997) 2662.

6, D.Bressanini, M.Mella, and G.Morosi, Phys. Rev. A55 (1997) 200

7, K.Varga, J.Usukura, and Y. Suzuki, Phys. Rev. Lett. 80 (1998) 1876

8, J.Usukura, K.Varga, and Y. Suzuki, Physics/9804023

9, C.G.Bao, in press in Phys. Lett. A.

10, C.G.Bao, Chin. Phys. Lett. 14 (1997) 20; Commu. Theor. Phys. 28 (1997)
363; Phys. Rev. Lett. 79 (1997) 3475.

\hspace{1.0in}

\hspace{1.0in}

\begin{tabular}{|c|c|c|c|c|c|c|c|c|c|c|c|}
\hline
L$^\Pi $ & $\mu $ & SQ & CH & DIA & REC & REC' & TE & TE' & Ps-Ps & 
Ps-Ps*(10) & Ps-Ps*(01) \\ \hline
0$^{+}$ & A$_1$ & a & a & a & a & a & a & a & open & open &  \\ \hline
0$^{+}$ & B$_2$ &  &  & a &  &  &  &  &  &  & open \\ \hline
0$^{+}$ & B$_1$ &  & a &  & a & a &  & a & open & open &  \\ \hline
0$^{+}$ & A$_2$ &  &  &  &  &  &  &  &  &  & open \\ \hline
0$^{+}$ & E &  & a &  &  &  &  &  &  & open & open \\ \hline
0$^{-}$ & A$_1$ &  &  &  &  &  &  & a &  &  &  \\ \hline
0$^{-}$ & B$_2$ &  &  &  &  &  &  &  &  &  &  \\ \hline
0$^{-}$ & B$_1$ &  &  &  &  &  & a & a &  &  &  \\ \hline
0$^{-}$ & A$_2$ &  &  &  &  &  &  &  &  &  &  \\ \hline
0$^{-}$ & E &  &  &  &  &  &  &  &  &  &  \\ \hline
\end{tabular}

Table 1a

\begin{tabular}{|cccccccccccc|}
\hline
\multicolumn{1}{|c|}{L$^\Pi $} & \multicolumn{1}{c|}{$\mu $} & 
\multicolumn{1}{c|}{SQ} & \multicolumn{1}{c|}{CH} & \multicolumn{1}{c|}{DIA}
& \multicolumn{1}{c|}{REC} & \multicolumn{1}{c|}{REC'} & \multicolumn{1}{c|}{
TE} & \multicolumn{1}{c|}{TE'} & \multicolumn{1}{c|}{Ps-Ps} & 
\multicolumn{1}{c|}{Ps-Ps*(10)} & Ps-Ps*(01) \\ \hline
\multicolumn{1}{|c|}{1$^{+}$} & \multicolumn{1}{c|}{A$_1$} & 
\multicolumn{1}{c|}{} & \multicolumn{1}{c|}{} & \multicolumn{1}{c|}{} & 
\multicolumn{1}{c|}{} & \multicolumn{1}{c|}{} & \multicolumn{1}{c|}{} & 
\multicolumn{1}{c|}{} & \multicolumn{1}{c|}{} & \multicolumn{1}{c|}{} &  \\ 
\hline
\multicolumn{1}{|c|}{1$^{+}$} & \multicolumn{1}{c|}{B$_2$} & 
\multicolumn{1}{c|}{} & \multicolumn{1}{c|}{} & \multicolumn{1}{c|}{} & 
\multicolumn{1}{c|}{a} & \multicolumn{1}{c|}{} & \multicolumn{1}{c|}{} & 
\multicolumn{1}{c|}{a} & \multicolumn{1}{c|}{} & \multicolumn{1}{c|}{} & open
\\ \hline
\multicolumn{1}{|c|}{1$^{+}$} & \multicolumn{1}{c|}{B$_1$} & 
\multicolumn{1}{c|}{} & \multicolumn{1}{c|}{} & \multicolumn{1}{c|}{a} & 
\multicolumn{1}{c|}{} & \multicolumn{1}{c|}{} & \multicolumn{1}{c|}{} & 
\multicolumn{1}{c|}{} & \multicolumn{1}{c|}{} & \multicolumn{1}{c|}{} &  \\ 
\hline
\multicolumn{1}{|c|}{1$^{+}$} & \multicolumn{1}{c|}{A$_2$} & 
\multicolumn{1}{c|}{a} & \multicolumn{1}{c|}{} & \multicolumn{1}{c|}{a} & 
\multicolumn{1}{c|}{a} & \multicolumn{1}{c|}{} & \multicolumn{1}{c|}{a} & 
\multicolumn{1}{c|}{a} & \multicolumn{1}{c|}{} & \multicolumn{1}{c|}{} & open
\\ \hline
\multicolumn{1}{|c|}{1$^{+}$} & \multicolumn{1}{c|}{E} & \multicolumn{1}{c|}{
} & \multicolumn{1}{c|}{} & \multicolumn{1}{c|}{} & \multicolumn{1}{c|}{} & 
\multicolumn{1}{c|}{a} & \multicolumn{1}{c|}{a} & \multicolumn{1}{c|}{a} & 
\multicolumn{1}{c|}{} & \multicolumn{1}{c|}{} & open \\ \hline
\multicolumn{1}{|c|}{1$^{-}$} & \multicolumn{1}{c|}{A$_1$} & 
\multicolumn{1}{c|}{} & \multicolumn{1}{c|}{} & \multicolumn{1}{c|}{} & 
\multicolumn{1}{c|}{} & \multicolumn{1}{c|}{} & \multicolumn{1}{c|}{} & 
\multicolumn{1}{c|}{} & \multicolumn{1}{c|}{} & \multicolumn{1}{c|}{open} & 
\\ \hline
\multicolumn{1}{|c|}{1$^{-}$} & \multicolumn{1}{c|}{B$_2$} & 
\multicolumn{1}{c|}{} & \multicolumn{1}{c|}{a} & \multicolumn{1}{c|}{} & 
\multicolumn{1}{c|}{} & \multicolumn{1}{c|}{a} & \multicolumn{1}{c|}{a} & 
\multicolumn{1}{c|}{a} & \multicolumn{1}{c|}{} & \multicolumn{1}{c|}{} & open
\\ \hline
\multicolumn{1}{|c|}{1$^{-}$} & \multicolumn{1}{c|}{B$_1$} & 
\multicolumn{1}{c|}{} & \multicolumn{1}{c|}{} & \multicolumn{1}{c|}{} & 
\multicolumn{1}{c|}{} & \multicolumn{1}{c|}{} & \multicolumn{1}{c|}{} & 
\multicolumn{1}{c|}{} & \multicolumn{1}{c|}{} & \multicolumn{1}{c|}{open} & 
\\ \hline
\multicolumn{1}{|c|}{1$^{-}$} & \multicolumn{1}{c|}{A$_2$} & 
\multicolumn{1}{c|}{} & \multicolumn{1}{c|}{a} & \multicolumn{1}{c|}{} & 
\multicolumn{1}{c|}{} & \multicolumn{1}{c|}{a} & \multicolumn{1}{c|}{} & 
\multicolumn{1}{c|}{a} & \multicolumn{1}{c|}{} & \multicolumn{1}{c|}{} & open
\\ \hline
\multicolumn{1}{|c|}{1$^{-}$} & \multicolumn{1}{c|}{E} & \multicolumn{1}{c|}{
a} & \multicolumn{1}{c|}{a} & \multicolumn{1}{c|}{a} & \multicolumn{1}{c|}{a}
& \multicolumn{1}{c|}{a} & \multicolumn{1}{c|}{a} & \multicolumn{1}{c|}{a} & 
\multicolumn{1}{c|}{open} & \multicolumn{1}{c|}{open} & open \\ \hline
\end{tabular}

Table 1b

\begin{tabular}{|cccccccccccc|}
\hline
\multicolumn{1}{|c|}{L$^\Pi $} & \multicolumn{1}{c|}{$\mu $} & 
\multicolumn{1}{c|}{SQ} & \multicolumn{1}{c|}{CH} & \multicolumn{1}{c|}{DIA}
& \multicolumn{1}{c|}{REC} & \multicolumn{1}{c|}{REC'} & \multicolumn{1}{c|}{
TE} & \multicolumn{1}{c|}{TE'} & \multicolumn{1}{c|}{Ps-Ps} & 
\multicolumn{1}{c|}{Ps-Ps*(10)} & Ps-Ps*(01) \\ \hline
\multicolumn{1}{|c|}{2$^{+}$} & \multicolumn{1}{c|}{A$_1$} & 
\multicolumn{1}{c|}{a} & \multicolumn{1}{c|}{a} & \multicolumn{1}{c|}{a} & 
\multicolumn{1}{c|}{a} & \multicolumn{1}{c|}{a} & \multicolumn{1}{c|}{a} & 
\multicolumn{1}{c|}{a} & \multicolumn{1}{c|}{open} & \multicolumn{1}{c|}{open
} &  \\ \hline
\multicolumn{1}{|c|}{2$^{+}$} & \multicolumn{1}{c|}{B$_2$} & 
\multicolumn{1}{c|}{a} & \multicolumn{1}{c|}{} & \multicolumn{1}{c|}{a} & 
\multicolumn{1}{c|}{a} & \multicolumn{1}{c|}{} & \multicolumn{1}{c|}{a} & 
\multicolumn{1}{c|}{a} & \multicolumn{1}{c|}{} & \multicolumn{1}{c|}{} & open
\\ \hline
\multicolumn{1}{|c|}{2$^{+}$} & \multicolumn{1}{c|}{B$_1$} & 
\multicolumn{1}{c|}{a} & \multicolumn{1}{c|}{a} & \multicolumn{1}{c|}{a} & 
\multicolumn{1}{c|}{a} & \multicolumn{1}{c|}{a} & \multicolumn{1}{c|}{a} & 
\multicolumn{1}{c|}{a} & \multicolumn{1}{c|}{open} & \multicolumn{1}{c|}{open
} &  \\ \hline
\multicolumn{1}{|c|}{2$^{+}$} & \multicolumn{1}{c|}{A$_2$} & 
\multicolumn{1}{c|}{} & \multicolumn{1}{c|}{} & \multicolumn{1}{c|}{a} & 
\multicolumn{1}{c|}{a} & \multicolumn{1}{c|}{} & \multicolumn{1}{c|}{} & 
\multicolumn{1}{c|}{a} & \multicolumn{1}{c|}{} & \multicolumn{1}{c|}{} & open
\\ \hline
\multicolumn{1}{|c|}{2$^{+}$} & \multicolumn{1}{c|}{E} & \multicolumn{1}{c|}{
} & \multicolumn{1}{c|}{a} & \multicolumn{1}{c|}{} & \multicolumn{1}{c|}{} & 
\multicolumn{1}{c|}{a} & \multicolumn{1}{c|}{a} & \multicolumn{1}{c|}{a} & 
\multicolumn{1}{c|}{} & \multicolumn{1}{c|}{open} & open \\ \hline
\multicolumn{1}{|c|}{2$^{-}$} & \multicolumn{1}{c|}{A$_1$} & 
\multicolumn{1}{c|}{} & \multicolumn{1}{c|}{} & \multicolumn{1}{c|}{} & 
\multicolumn{1}{c|}{} & \multicolumn{1}{c|}{} & \multicolumn{1}{c|}{a} & 
\multicolumn{1}{c|}{a} & \multicolumn{1}{c|}{} & \multicolumn{1}{c|}{} &  \\ 
\hline
\multicolumn{1}{|c|}{2$^{-}$} & \multicolumn{1}{c|}{B$_2$} & 
\multicolumn{1}{c|}{} & \multicolumn{1}{c|}{} & \multicolumn{1}{c|}{} & 
\multicolumn{1}{c|}{} & \multicolumn{1}{c|}{a} & \multicolumn{1}{c|}{} & 
\multicolumn{1}{c|}{a} & \multicolumn{1}{c|}{} & \multicolumn{1}{c|}{} & open
\\ \hline
\multicolumn{1}{|c|}{2$^{-}$} & \multicolumn{1}{c|}{B$_1$} & 
\multicolumn{1}{c|}{} & \multicolumn{1}{c|}{} & \multicolumn{1}{c|}{} & 
\multicolumn{1}{c|}{} & \multicolumn{1}{c|}{} & \multicolumn{1}{c|}{a} & 
\multicolumn{1}{c|}{a} & \multicolumn{1}{c|}{} & \multicolumn{1}{c|}{} &  \\ 
\hline
\multicolumn{1}{|c|}{2$^{-}$} & \multicolumn{1}{c|}{A$_2$} & 
\multicolumn{1}{c|}{} & \multicolumn{1}{c|}{} & \multicolumn{1}{c|}{} & 
\multicolumn{1}{c|}{} & \multicolumn{1}{c|}{a} & \multicolumn{1}{c|}{a} & 
\multicolumn{1}{c|}{a} & \multicolumn{1}{c|}{} & \multicolumn{1}{c|}{} & open
\\ \hline
\multicolumn{1}{|c|}{2$^{-}$} & \multicolumn{1}{c|}{E} & \multicolumn{1}{c|}{
a} & \multicolumn{1}{c|}{} & \multicolumn{1}{c|}{a} & \multicolumn{1}{c|}{a}
& \multicolumn{1}{c|}{a} & \multicolumn{1}{c|}{a} & \multicolumn{1}{c|}{a} & 
\multicolumn{1}{c|}{} & \multicolumn{1}{c|}{} & open \\ \hline
\end{tabular}

Table 1c

Table 1a-1c The accessibility of the regular shapes and the accessibility of
the Ps-Ps and Ps-Ps* outgoing channels. The SQ denotes a square with the two 
$e^{-}(e^{+})$ located at opposite vertexes. The CH denotes a straight
chain, the particles lying along the chain are symmetric to the c.m., and
adjacent particles havs opposite charges. The DIA\ denotes a diamond with
the two $e^{-}(e^{+})$ located at opposite vertexes. The REC\ (REC') denotes
a rectangle with the two $e^{-}(e^{+})$ located at opposite (adjacent)
vertexes. The TE\ (TE') denotes a regular tetrahedron which may be prolonged
or flattened along one of its 2-fold axis and with each pair of particles
being symmetric to the axis having the same (opposite) charge(s). A block
with a ''a'' (''open'') implies that the associated L$^\Pi (\mu )$ states
can access the associated shape (channel), otherwise it can not. In the last
two columns the (01) and (10) are the ($nl$) of the excited positronium.

\hspace{1.0in}

\hspace{1.0in}

\begin{tabular}{|c|c|c|c|c|}
\hline
L & Bound states & Bound states & Bound states & Resonances \\ 
& predicted by theory & anticipated by symmetry & or resonances & from Ps-Ps
channel \\ \hline
0 & 0$^{+}(A_1),$0$^{+}(B_2),$0$^{+}(E)$ & 0$^{-}(\mu )$ & 0$^{+}(B_1),$0$%
_2^{+}(A_1),$0$^{+}(A_2)$ & 0$_2^{+}(B_1),$0$_3^{+}(A_1),\cdot \cdot \cdot
\cdot $ \\ \hline
1 & 1$^{-}(B_2)$ & 1$^{+}(A_2),$1$^{-}(A_2),$ & 1$^{-}(E)$ & 1$%
_2^{-}(E),\cdot \cdot \cdot \cdot $ \\ 
&  & 1$^{+}(A_1),$1$^{+}(B_1)$ &  &  \\ \hline
2 &  & 2$^{-}(A_1),$2$^{-}(B_1)$ & 2$^{+}(B_2),$2$^{+}(E),$2$^{-}(E)$ & 2$%
_{}^{+}(A_1),$2$_{}^{+}(B_1),\cdot \cdot \cdot \cdot $ \\ \hline
\end{tabular}

Table 2 Bound states and resonances anticipated by theory calculation or by
symmetry consideration.

\end{document}